\documentclass[aps,pre,twocolumn,superscriptaddress,showpacs]{revtex4-1}
\usepackage{graphicx}
\usepackage{array}
\usepackage{amsfonts}
\usepackage{amssymb,amsmath,multirow,rotate}
\usepackage{mathrsfs}
\usepackage{booktabs}
\usepackage{threeparttable}
\usepackage{multirow}
\usepackage{epsfig}
\usepackage{threeparttable}
\usepackage{chngpage}
\usepackage{latexsym}
\usepackage{dcolumn}
\usepackage{graphics,graphicx,bm,fleqn,epic,eepic,float}
\usepackage{verbatim}
\usepackage{subfigure}
\usepackage{color}
\usepackage[english]{babel}

\definecolor{red}{rgb}{1,0,0}

\definecolor{blue}{rgb}{0,0,1}

\begin{document}
%\draft

\title{Transition to high-dimensional chaos in nonsmooth dynamical systems}

\date{\today}

\author{Ru-Hai Du}
\affiliation{School of Physics and Information Technology, Shaanxi Normal University, Xi'an 710062, China}

\author{Shi-Xian Qu} \email{Corresponding author: sxqu@snnu.edu.cn}
\affiliation{School of Physics and Information Technology, Shaanxi Normal University, Xi'an 710062, China}

\author{Ying-Cheng Lai}
\affiliation{School of Physics and Information Technology, Shaanxi Normal University, Xi'an 710062, China}
\affiliation{School of Electrical, Computer and Energy Engineering, Arizona State University, Tempe, AZ 85287, USA}
\affiliation{Department of Physics, Arizona State University,
Tempe, Arizona 85287, USA}

\begin{abstract}

We uncover a route from low-dimensional to high-dimensional chaos in nonsmooth
dynamical systems as a bifurcation parameter is continuously varied. The
striking feature is the existence of a finite parameter interval of periodic
attractors in between the regimes of low- and high-dimensional chaos. That is,
the emergence of high-dimensional chaos is preceded by the system's settling
into a totally nonchaotic regime. This is characteristically distinct from
the situation in smooth dynamical systems where high-dimensional chaos emerges
directly and smoothly from low-dimensional chaos. We carry out an analysis to
elucidate the underlying mechanism for the abrupt emergence and disappearance
of the periodic attractors and provide strong numerical support for the
typicality of the transition route in the pertinent two-dimensional parameter
space. The finding has implications to applications where high-dimensional
and robust chaos is desired.

\end{abstract}
\maketitle

%\tableofcontents

%\newpage
\section{Introduction} \label{sec:intro}

In nonlinear dynamical systems, there are two kinds of chaos: low-dimensional
and high-dimensional. The characteristic feature of a low-dimensional
chaotic invariant set (e.g., an attractor) is that it has only one positive
Lyapunov exponent, examples of which include the classic
Lorenz~\cite{Lorenz:1963} and R\"{o}ssler~\cite{Rossler:1976} attractors.
High-dimensional chaotic sets are those that possess more than one positive
Lyapunov exponent~\cite{IM:1987,BHGY:1997,MM:2009,IMAD:2015}.
The difference between low- and high-dimensional chaos
can be appreciated in terms of significant issues such as control. The
well established Ott-Grebogi-Yorke
paradigm~\cite{OGY:1990,BGLMM:2000} of controlling chaos
is most effective for low-dimensional chaotic systems, and controlling
high-dimensional chaos~\cite{AGOY:1992,GL:1997,BGLMM:2000} has remained
to be a challenging problem.

In nonlinear dynamics, the four routes to chaos, namely,
period-doubling~\cite{Feigenbaum:1978}, intermittency~\cite{PM:1980},
crisis~\cite{GOY:1983}, and
quasiperiodicity~\cite{RT:1971,GS:1975,NRT:1978},
concerned about the bifurcation to a low-dimensional chaotic attractor.
There was also work on the transition from low- to high-dimensional
chaos~\cite{HL:1999,HL:2000,DL:2000,PSM:2001,PM:2005}
in smooth dynamical systems. A general phenomenon is that
the second Lyapunov exponent passes through zero and becomes positive
smoothly as a parameter changes through the transition point. Heuristically,
this can be understood in light of transition to chaos in random dynamical
systems~\cite{YOC:1990,RHKK:2000,LLBS:2002,LLBS:2003,XLZD:2003}, in 
systems with a symmetry~\cite{Lai:1996a}, and in quasiperiodically driven 
systems~\cite{Lai:1996b,LFG:1996,YL:1996}.
In particular, before the transition,
the system is already chaotic with one positive Lyapunov exponent - the
largest exponent giving the exponential growth of an infinitesimal vector
in the corresponding tangent subspace. For convenience, we call it the
primary tangent subspace. The evolution of the vector in the secondary tangent
subspace corresponding to the second nontrivial Lyapunov exponent can then
be regarded as being subject to an indirect, chaotic or effectively ``random''
driving from the dynamics in the primary tangent subspace. Under such a
driving, an infinitesimal vector in the secondary tangent subspace will
exhibit temporal episodes of ``expansion'' or ``contraction,'' a generic
feature in random dynamical systems~\cite{YOC:1990}. The second nontrivial
Lyapunov exponent will then exhibit fluctuations in finite time between
positive and negative values. The asymptotic value of the exponent depends
on the relative weights of the underlying infinitesimal tangent
vector in the expanding or contracting phase~\cite{HL:1999}:
the value is negative if the contraction phase
over weighs the expansion phase, and vice versa for a positive value.
Transition to high-dimensional chaos occurs when the contraction and
expansion weights are balanced. Since the weights vary smoothly with
the bifurcation parameter~\cite{HL:1999}, the second nontrivial Lyapunov
exponent passes through zero smoothly. The generic feature associated
with the transition in smooth dynamical systems is thus that high-dimensional
chaos arises directly and smoothly from low-dimensional
chaos~\cite{HL:1999,HL:2000,DL:2000}.

\begin{figure}[htbp]
\centering
\includegraphics[width=\linewidth]{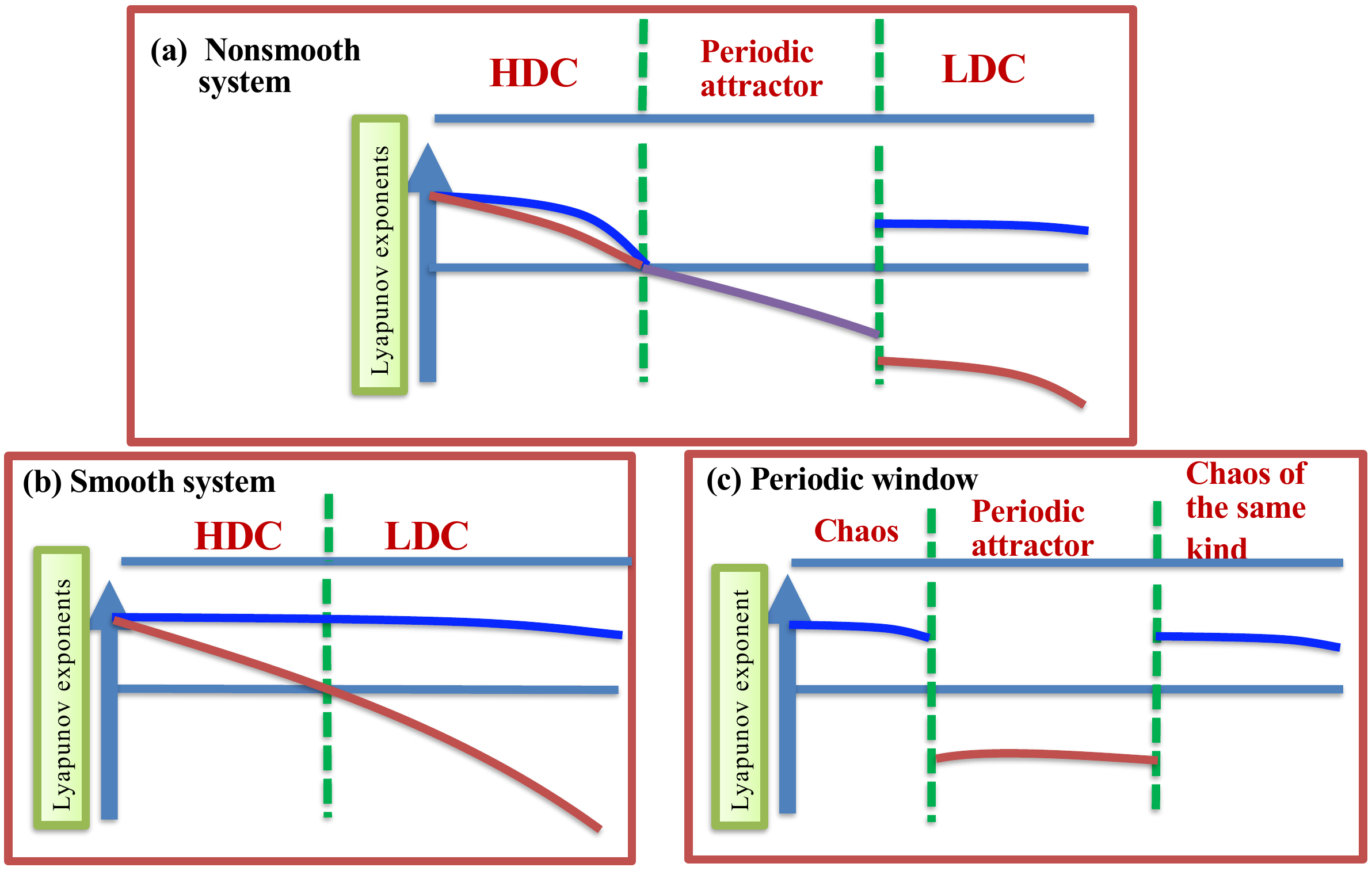}
\caption{ \textbf{Schematic illustration of transition scenarios to
high-dimensional chaos (HDC) in nonsmooth and smooth dynamical systems}.
(a) The main result of the paper. As a parameter (e.g., the coupling
strength) is reduced from the regime of low-dimensional chaos (LDC) with a
single positive Lyapunov exponent, an open parameter interval of periodic
attractors arises before HDC emerges. This transition
scenario should be contrasted to that in smooth dynamical systems in (b)
and is also fundamentally different from the dynamical behavior before
and after the occurrence of a periodic window in (c). (b) Transition to
HDC in smooth dynamical systems where the second largest Lyapunov exponent
passes through zero smoothly at the transition point. That is, HDC emerges
exactly where LDC ends. (c) The behavior about a periodic window, where chaos
on both sides of the window is of the same kind: either low-dimensional or
high-dimensional.}
\label{fig:schematic}
\end{figure}

In this paper, we investigate the transition to high-dimensional chaos in
nonsmooth dynamical systems that arise commonly in physical, engineering,
and biological applications such as impact oscillators~\cite{TG:1983,SH:1983,
Whiston:1983,Nordmark:1991,CONG:1994,CCGO:1996},
electronic circuits~\cite{BYG:1998,BG:1999,BRG:2000},
and neuronal networks~\cite{NC:2016,NTS:2016}. Mathematically, a typical
representation of such systems is piecewise smooth systems, e.g.,
a one-dimensional piecewise smooth map that can generate low-dimensional
chaos. For a system with two pieces, the phase space can be divided into
two regions where the dynamical equations in each region are different
but are nevertheless smooth, with a border separating the two regions.
This setting is representative of physical systems such as electronic
switching circuits~\cite{BYG:1998,BG:1999,BRG:2000}.
Previous mathematical analyses of piecewise
smooth systems with low-dimensional chaos revealed interesting phenomena
such as period-adding bifurcations and transition to chaos from a periodic
attractor of arbitrary period, as a result of ``border collision'' in
phase space~\cite{NY:1992,NOY:1994,DNOYY:1999,LZCY:2002,HAN:2004,GB:2005,ASB:2007,DL:2008}.
Because our goal is to uncover and understand
how high-dimensional chaos may arise from low-dimensional chaos in
nonsmooth systems, we consider the minimal setting of two coupled piecewise
smooth subsystems, each capable of exhibiting low-dimensional chaos. At zero
coupling, the two subsystems are isolated and the system as a whole has
two positive Lyapunov exponents - a trivial type of high-dimensional
chaos. In the weak coupling regime, there is interaction between the two
subsystems and the system possesses high-dimensional chaos. In the strong
coupling regime, synchronization between the two subsystems can occur, and
the dynamics of the full system are effectively those of a single subsystem.
As a result, the system exhibits low-dimensional chaos. In the intermediate
coupling regime, a transition between low- and high-dimensional chaos can be
expected. Is the transition scenario any different than that in smooth
dynamical systems?

The main finding of this paper is that, in nonsmooth dynamical systems,
there can be two distinct routes to high-dimensional chaos: one that is
similar to and another characteristically different from that in smooth
dynamical systems. In particular, depending on the system parameter values,
high-dimensional chaos can arise directly and smoothly from low-dimensional
chaos, as in smooth dynamical systems. The striking phenomenon is the existence
of an open parameter region in nonsmooth systems where a nonchaotic, ``buffer''
regime with a periodic attractor arises in between regions of low- and
high-dimensional chaos. For example, for the minimal coupled nonsmooth system,
as the coupling parameter is decreased from the low-dimensional, synchronous
chaos regime, a periodic attractor can arise abruptly and last for a finite
parameter interval. At a smaller parameter value, a high-dimensional chaotic
attractor emerges abruptly from the periodic attractor. The ``buffer''
periodic attractor occurs in an open interval of the coupling parameter.
In a two-dimensional parameter space, the buffer or ``precursor'' periodic
attractor occupies a finite region - a ``bubble,'' signifying its typicality.
The emergence of the bubble region can generally be attributed to border
collision bifurcations, for which we provide a detailed analysis. The same
transition scenario can occur when the phase space dimension is much larger
than two, e.g., in a system of a large number of coupled nonsmooth maps.
In such a case, high-dimensional chaos manifests itself as synchronous
clusters with distinct chaotic behaviors, low-dimensional chaos corresponds
to globally synchronous chaos, and periodic synchronization occurs in the
bubble region. A schematic illustration of our main result and its
characteristic difference from the transition scenario in smooth dynamical
systems as well as from that around a periodic window is presented in
Fig.~\ref{fig:schematic}.

\section{Model and results} \label{sec:result}

\subsection{A system of coupled piecewise linear maps and Lyapunov exponents}
\label{subsec:model}

A typical class of nonsmooth dynamical systems is piecewise smooth
maps~\cite{CHHBKM:1992,QWH:1998,WDHWMH:2001,LHJ:2005,YWQ:2015}.
To investigate the transition route to high-dimensional chaos, we use
coupled map lattices~\cite{Kaneko:1989,Kaneko:1990a,Kaneko:1990b,PPM:2001,
PMM:2001,PMM:2002,TKWTBT:2008,PBH:2009,PR:2015,Kaneko:2015}. Specifically,
we consider the following system of $N$ globally coupled, piecewise smooth
maps:
\begin{equation} \label{eq:GCM}
x_{n+1}(i)=(1-\varepsilon)f[x_n(i)]+\frac{\varepsilon}{N}\sum_{j=1}^Nf[x_n(j)],
\end{equation}
where $x_n(i)$ is the dynamical variable of the $i$th node at time $n$,
$\varepsilon\in[0,1]$ is a coupling parameter, and $f(x)$ represents the
nodal dynamics. To be concrete, we consider the following one-dimensional
piecewise linear map
\begin{equation}\label{map}
f(x)= \left\{ \begin{array}{ll}
\alpha x-\mu, & ~x<0,\\
\beta x-\mu-\gamma, ~&~ x>0,
\end{array}
\right.
\end{equation}
where $\alpha, \beta, \mu$ and $\gamma$ are parameters. There is a nonsmooth
border at $x=0$ where the left and right limits of the mapping function are
not identical. To investigate the transition to
high-dimensional chaos, we require that the isolated nodal dynamics generate
a low-dimensional chaotic attractor, which can be realized for, e.g., the
following parameter setting: $\alpha=0.4, \beta=-8, \mu=0.1$, and $\gamma=0$.
The Lyapunov exponent of this attractor is $\lambda_0 \approx 0.251$.

A typical dynamical state of system (\ref{eq:GCM}) is
cluster formation, where the dynamics of all nodes within a cluster are
synchronized but those among different clusters are unsynchronized.
For simplicity, we consider the case of a two-cluster state~\cite{PBH:2009}:
\begin{equation} \label{eq:2clusters}
\begin{array}{ll}
x_n(1) = x_n(2) = \cdot\cdot\cdot = x_n(N_1) = x_n, \\
x_n(N_1+1) = x_n(N_1+2) = \cdot\cdot\cdot = x_n(N) = y_n,
\end{array}
\end{equation}
where $N_1$ and $N_2=N-N_1$ are the numbers of nodes in the two clusters.
Because of synchronization within each cluster, we obtain an effective
two-dimensional nonsmooth map:
\begin{eqnarray} \label{eq:2DCM}
x_{n+1} & = & [1 - \varepsilon(1-r))f(x_n)+\varepsilon(1-r)f(y_n), \\ \nonumber
y_{n+1} & = & \varepsilon rf(x_n)+(1-\varepsilon r)f(y_n),
\end{eqnarray}
where $r=N_1/N$ is the fraction of nodes belonging to the $x$ cluster.
In the thermodynamic limit $N\to\infty$, $r$ is a continuous parameter.
System~(\ref{eq:2DCM}) describes the dynamical evolution of the two cluster
state in system~(\ref{eq:GCM}). For $r=1/2$, the two clusters are symmetric
with respect to each other~\cite{PBH:2009}. As we will demonstrate, the two
distinct routes to high-dimensional chaos occur in different intervals of
$r$ values.

Because of the mirror symmetry with respect to $r=1/2$, we introduce the
parameter $\bar{r}=r-1/2$ to rewrite Eq.~(\ref{eq:2DCM}) as
\begin{equation} \label{eq:N2DCM}
\begin{array}{ll}
x_{n+1}=(1-\varepsilon(\frac{1}{2}-\bar{r}))x_n+\varepsilon(\frac{1}{2}-\bar{r})y_n,\\
y_{n+1}=\varepsilon(\frac{1}{2}+\bar{r})x_n+(1-\varepsilon (\frac{1}{2}+\bar{r}))y_{n},
\end{array}
\end{equation}
which has an exact synchronous solution: $x_n=y_n=s_n$. We write
${\bf x}_{n+1} = {\bf F}({\bf x}_n)$, where ${\bf x} \equiv (x,y)^T$ and
the symbol ``$T$'' denotes transpose. The corresponding variational
equations are
\begin{equation}\label{eq:VE}
\!\left(\!\!
\begin{array}{l}
  \!\delta x_{n+1}\\
  \!\delta y_{n+1}\\
\end{array}\!
\!\!\!\right)\!\!=\!\!f'\!(\!s_n\!)\!
\!\left(\!\!\!
  \begin{array}{cc}
    1-\varepsilon(\frac{1}{2}-\bar{r}) & \varepsilon(\frac{1}{2}-\bar{r}) \\
    \varepsilon (\frac{1}{2}+\bar{r}) & 1-\varepsilon (\frac{1}{2}+\bar{r}) \\
  \end{array}\!\!
\!\right)\!\!
\!\left(\!
  \!\begin{array}{l}
    \!\delta x_n \\
    \!\delta y_n \\
  \end{array}\!
\!\!\right),
\end{equation}
where $\delta x_n=x_n-s_n$, $\delta y_n=y_n-s_n$, and $f'(s_n)$ is the
derivative of the map function evaluated at the synchronization manifold.
The two eigenvalues of the coupling matrix are $u_1=1$ and $u_2=1-\varepsilon$.
The corresponding transform matrix is given by
\begin{equation}
\mathbf{Q}=\left(
  \begin{array}{cc}
    1 & \frac{\bar{r}-\frac{1}{2}}{\frac{1}{2}+\bar{r}} \\
    1 & 1 \\
  \end{array}
\right).
\end{equation}
The transform $(\delta\tilde{x}_n,\delta\tilde{y}_n)^T
= \mathbf{Q}^{-1}\cdot(\delta x_n, \delta y_n)^T$ leads to a diagonally
decoupled form of Eq.~(\ref{eq:VE}):
\begin{equation}
\left(\!
  \begin{array}{c}
    \delta \tilde{x}_{n+1} \\
    \delta \tilde{y}_{n+1} \\
  \end{array}\!
\right)
=f'(s_n)\!\!
\left(\!
  \begin{array}{cc}
    1 & 0 \\
    0 & 1-\varepsilon \\
  \end{array}\!
\right)\!\!\!
\left(\!
  \begin{array}{c}
    \delta \tilde{x}_n \\
    \delta \tilde{y}_n \\
  \end{array}\!
\right)
\end{equation}
The transverse Lyapunov exponent is given by
\begin{equation}
\lambda_\perp=\ln(1-\varepsilon)+\lambda_0
\end{equation}
where $\lambda_0 > 0$ is the Lyapunov exponent of the chaotic attractor of the
individual map. Stable synchronization can be achieved for $\lambda_\perp <0$.
The critical value of the coupling parameter above which synchronization
occurs is $\varepsilon_c = 1 - e^{-\lambda_0} \approx 0.222$.

The Lyapunov exponents of the asymptotic invariant set of the system can be
calculated from the Jacobian matrix $\mathbf{DF}$ associated with a typical
trajectory $\{(x_n, y_n)\}_{n=0}^\infty$:
\begin{equation} \label{eq:Jacobian}
\mathbf{DF}^n(x_0, y_0)=\prod_{j=0}^{n-1}\mathbf{DF}(x_j, y_j).
\end{equation}
The eigenvalues of the Jacobian matrix are given by
\begin{equation}
\mathrm{det}(\mathbf{DF}^n-u\mathbf{I}) = 0.
\end{equation}
We have that the eigenvalue $u$ satisfies
\begin{equation}
u^2 -\tau u + \Delta = 0,
\end{equation}
where $\tau \equiv \mathrm{trace}(\mathbf{DF}^n)$ and
$\Delta \equiv \det(\mathbf{DF}^n)$. For $\tau^2-4\Delta \geq 0$, the
eigenvalues are real and the Lyapunov exponents are given by
\begin{equation}\label{eq:RealLE}
\lambda_{1}=\frac{1}{n}\ln|u_{1}|,~\
\lambda_{2}=\frac{1}{n}\ln|u_{2}|,
\end{equation}
where $u_{1}=(\tau+\sqrt{\tau^2-4\Delta})/2$ and
$u_{2}=(\tau-\sqrt{\tau^2-4\Delta})/2$. For $\tau^2-4\Delta<0$, we obtain
a pair of complex conjugate eigenvalues. In this case,
the Lyapunov exponents are determined by the absolute value of the
eigenvalues $|u|$. We have
\begin{equation}\label{ComplexLE}
\lambda_{1,2}=\frac{1}{n}\ln|\mathrm{Re}(u)|.
\end{equation}
where $\mathrm{Re}(u)=\Delta$. For a periodic attractor of period-m, we have
$n=m$.

\subsection{Main result: coexistence of distinct transition routes to
high-dimensional chaos} \label{subsec:result}

\begin{figure}[t]
\centering
\includegraphics[width=1.0\linewidth]{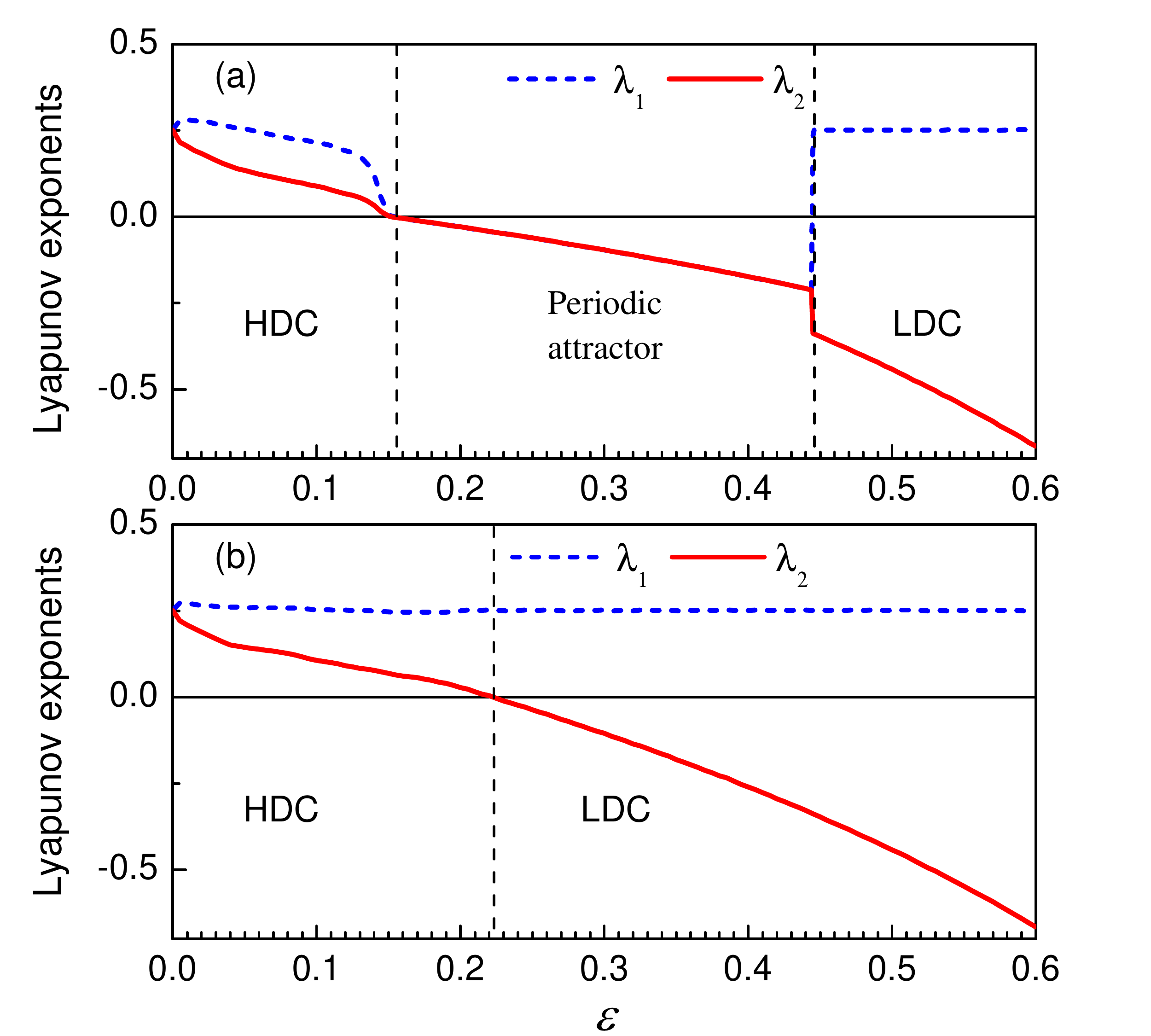}
\setlength{\abovecaptionskip}{0pt}
\setlength{\belowcaptionskip}{0pt}
\caption{(Color online)
{\bf Two distinct routes of transition to high-dimensional chaos in
nonsmooth dynamical systems}. (a) For $\bar{r}=0.3$, the transition route
follows the scenario in Fig.~\ref{fig:schematic}(a). As the coupling parameter
$\varepsilon$ is decreased from a relatively large value where there is
synchronous chaos with one positive Lyapunov exponent (LDC), a periodic
attractor arises. A high-dimensional chaotic attractor with two positive
Lyapunov exponents (HDC) emerges when the periodic attractor disappears.
(b) For $\bar{r}=0.48$, HDC arises directly and smoothly from LDC as in
smooth dynamical systems.}
\label{fig:Lypunovexp}
\end{figure}

For system~(\ref{eq:N2DCM}), we uncover a distinct route from low-dimensional
to high-dimensional chaos as the coupling parameter $\varepsilon$ is reduced.
In particular, for relatively large values of $\varepsilon$, there is
synchronous chaos and the system has a low-dimensional chaotic attractor
with one positive Lyapunov exponent. As $\varepsilon$ is decreased, a
periodic attractor with two identical negative Lyapunov exponents arises
abruptly and lasts for a finite parameter interval. High-dimensional chaos
with two positive Lyapunov exponents emerges where the periodic attractor
disappears. That is, there exists a ``buffer'' region of some periodic
attractor in between low- and high-dimensional chaos. This transition scenario
to high-dimensional chaos, as exemplified by Fig.~\ref{fig:Lypunovexp}(a)
[schematically illustrated in Fig.~\ref{fig:schematic}(a)] for $\bar{r}=0.3$,
is unique for nonsmooth dynamical systems. For a different value of
parameter $\bar{r}$, the typical route to high-dimensional chaos in
smooth dynamical systems [schematically illustrated in
Fig.~\ref{fig:schematic}(b)] occurs, where a high-dimensional chaotic
attractor emerges directly and smoothly from a low-dimensional chaotic
attractor, as demonstrated in Fig.~\ref{fig:Lypunovexp}(b) for $\bar{r}=0.48$.
Nonsmooth dynamical systems thus exhibit richer transition scenarios to
high-dimensional chaos than smooth systems.

\begin{figure}[t]
\includegraphics[width=1.0\linewidth]{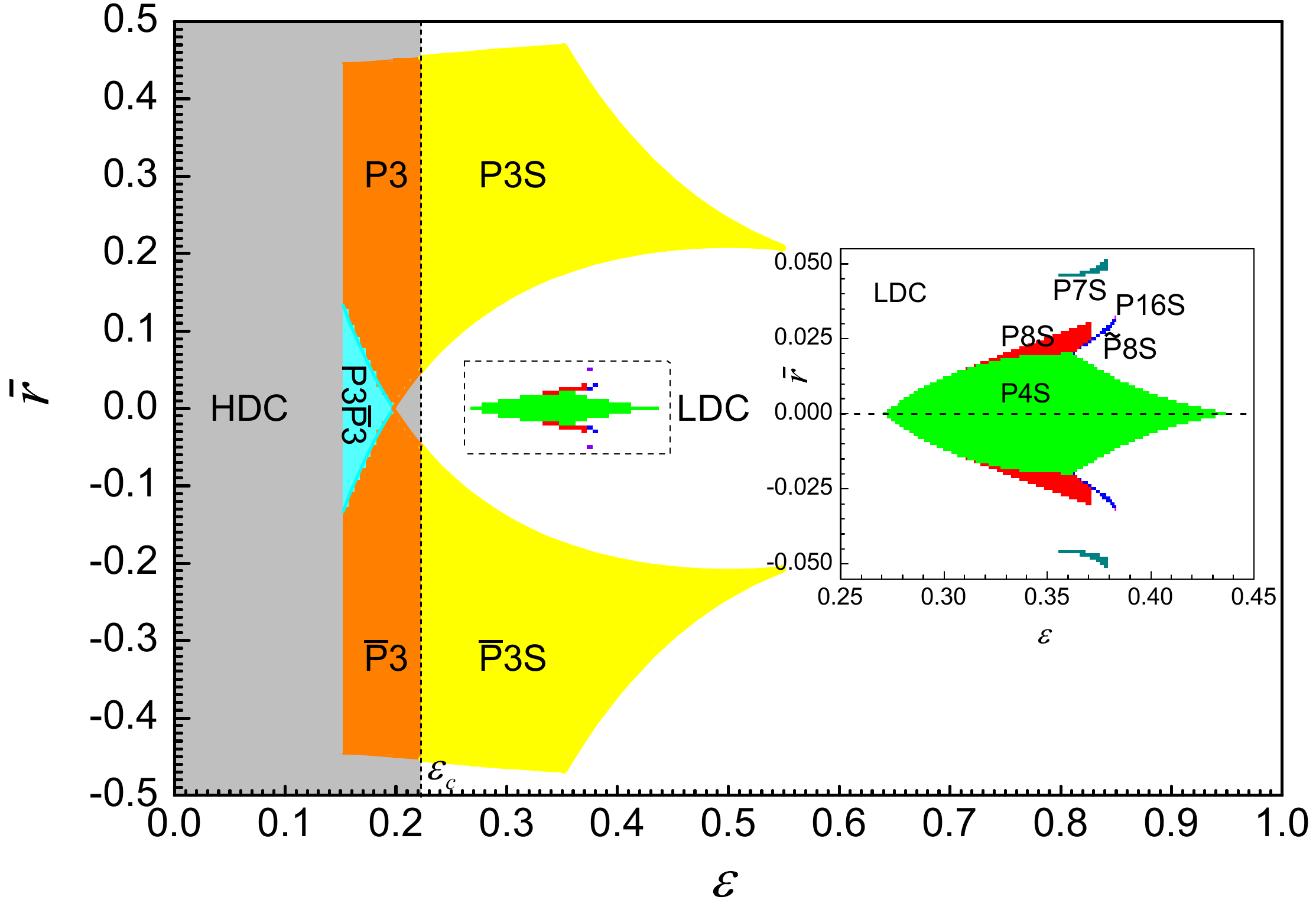}
\setlength{\abovecaptionskip}{0pt}
\setlength{\belowcaptionskip}{0pt}
\centering
\caption{(Color online)
{\bf Phase diagram of distinct attractors in the parameter plane
($\varepsilon$, $\bar{r}$)}. There is a mirror symmetry with
respect to $\bar{r}=0$. A period-m attractor is denoted as Pm. The vertical
dashed line represents the critical value $\varepsilon_c$ of the coupling
parameter beyond which synchronous chaos arises. Legends are: HDC -
high-dimensional chaos, LDC - low-dimensional chaos (fully synchronous
chaotic state - the blank region), P3 - clockwise period-3 attractor,
$\mathrm{\bar{P}}$3 - counterclockwise period-3 attractor. Additional legends
are: P3$\mathrm{\bar{P}}$3 - coexistence of two distinct period-3 attractors;
P3S: coexistence of a clockwise period-3 attractor with LDC;
$\mathrm{\bar{P}}$3S: counterclockwise period-3 attractor coexisting with
LDC. Inset is magnification of the region enclosed by the dashed rectangular
box in which periodic attractors of high periods (e.g., P8 and
$\mathrm{\tilde{P}}$8) coexist with LDC.}
\label{fig:phasediagram}
\end{figure}

To obtain a complete picture of the asymptotic attractors in different
parameter regions, we calculate the phase diagram in the parameter plane
$(\varepsilon,\bar{r})$, as shown in Fig.~\ref{fig:phasediagram}. The phase
diagram has a mirror symmetry about $\bar{r}=0$. For example, the parameter
regions in which two distinct period-3 attractors, P3 and $\mathrm{\bar{P}}$3,
occur are symmetric about $\bar{r}=0$. In the strongly coupling regime,
synchronous chaotic attractors arise: the system exhibits low-dimensional
chaos, whereas high-dimensional chaos occurs in the weakly coupling regime.
For $-0.45 \alt \bar{r} \alt 0.45$, the transition from low- to high-dimensional
chaos as the coupling parameter $\varepsilon$ is decreased follows
the route as demonstrated schematically in Fig.~\ref{fig:schematic}(a) and
realistically in Fig.~\ref{fig:Lypunovexp}(a), where the period-3
attractor occupies a large region in the parameter plane. For
$0.45 \alt |\bar{r}| < 0.5$, the transition from low- to high-dimensional chaos
follows the conventional route [Figs.~\ref{fig:schematic}(b) and
\ref{fig:Lypunovexp}(b)] as in smooth dynamical systems.
There are also regions in the parameter plane where periodic attractors
of various periods arise. For example, the region marked by
P3$\mathrm{\bar{P}}$3 is one in which two symmetric period-3 attractors
coexist, each with a distinct basin of attraction. There are also periodic
attractors as a result of period-doubling bifurcations, such as those
denoted as P4, P8, and P16, as well as those created by period-adding
bifurcations, e.g., P3, P4, and P7. In the following,
we carry out an analysis to elucidate the underlying mechanism for the
abrupt emergence of the periodic attractors in between regimes of low- and
high-dimensional chaos.

\section{Emergence of periodic attractors between regimes of low- and
high-dimensional chaos} \label{sec:analysis}

For nonsmooth dynamical system, linear stability analysis alone is often
inadequate to characterize the bifurcations or transitions~\cite{PBH:2009}.
We find that, in our piecewise linear systems, the transition from
low-dimensional chaos to a periodic attractor is typically of the second
order, continuous type. The dynamical origin of the transition is border
collision bifurcations.

\subsection{Emergence of period-3 attractors} \label{subsec:P3attractor}

\begin{figure}[t]
\centering
\includegraphics[width=1.0\linewidth]{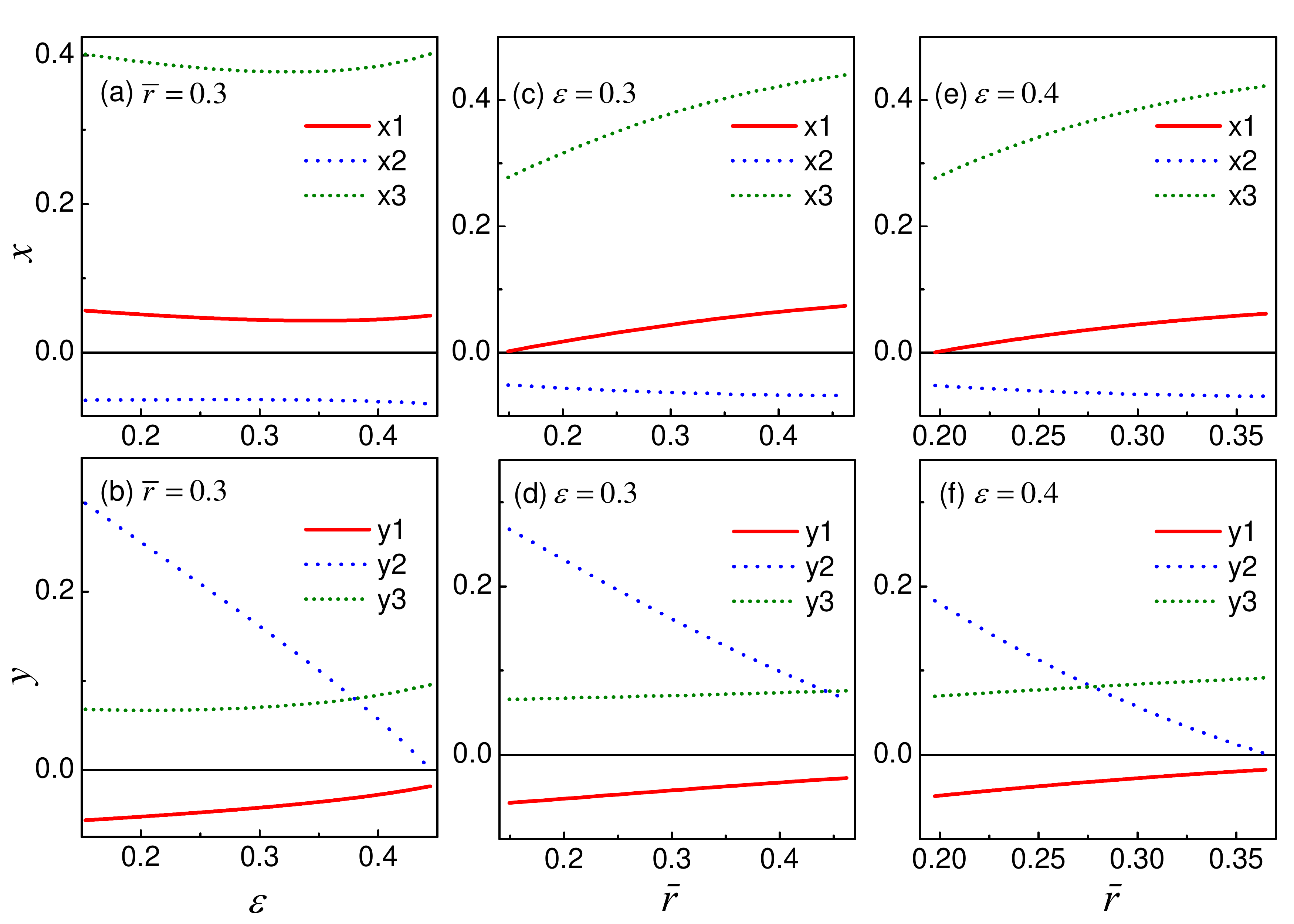}
\setlength{\abovecaptionskip}{0pt}
\setlength{\belowcaptionskip}{0pt}
\caption{(Color online)
{\bf Emergence and disappearance of a period-3 attractor}.
The three orbital points are denoted by $(x_1^*, y_1^*)$, $(x_2^*, y_2^*)$
and $(x_3^*, y_3^*)$. Shown are examples of how the orbital points of the
period-3 attractor depend on the bifurcation parameter $\varepsilon$ or
$\bar{r}$: (a,b) $\bar{r}=0.3$, (c,d) $\varepsilon=0.3$, and
(e,f) $\varepsilon=0.4$.}
\label{fig:P3orbits}
\end{figure}

The period-3 attractors take up a considerable region in the two-dimensional
parameter space. In order to determine the stability condition of the
attractor, we examine its orbital structure as a bifurcation parameter is
continuously varied. Taking advantage of the symmetry of the system, we
focus on the region of $\bar{r}\geq 0$. The three orbital points are denoted
as $(x_1^*, y_1^*)$, $(x_2^*, y_2^*)$ and $(x_3^*, y_3^*)$, where the first
point $(x_1^*, y_1^*)$ is located at the bottom of the phase portrait:
$y_1^*$ is the minimal value, as shown in Fig.~\ref{fig:P3orbits}.
For $\bar{r}=0.3$, from Figs.~\ref{fig:P3orbits}(a,b), we see that, at the
left boundary the stable period-3 attractor disappears without collision,
while at the right boundary it disappears because of the collision between
the $y_2$ orbit and the border $y=0$.
The phase space for $\varepsilon=0.3$ is shown in
Figs.~\ref{fig:P3orbits}(c,d), where the $x_1$ orbit collides with the border
$x=0$ for a small value of $\bar{r}$ and the stable period-3 attractor
disappears without collision for a large value of $\bar{r}$.
For $\varepsilon=0.45$, as shown in Figs.~\ref{fig:P3orbits}(e,f), we see that
the two boundaries of the disappearance of the period-3 attractor are both
due to border collision bifurcations. In particular, the $x_1$ orbit
collides with the border $x=0$ for a small value of $\bar{r}$ and the $y_2$
orbit collides with the border $y=0$ for a large value of $\bar{r}$.

\begin{figure}[t]
\centering
\includegraphics[width=1.0\linewidth]{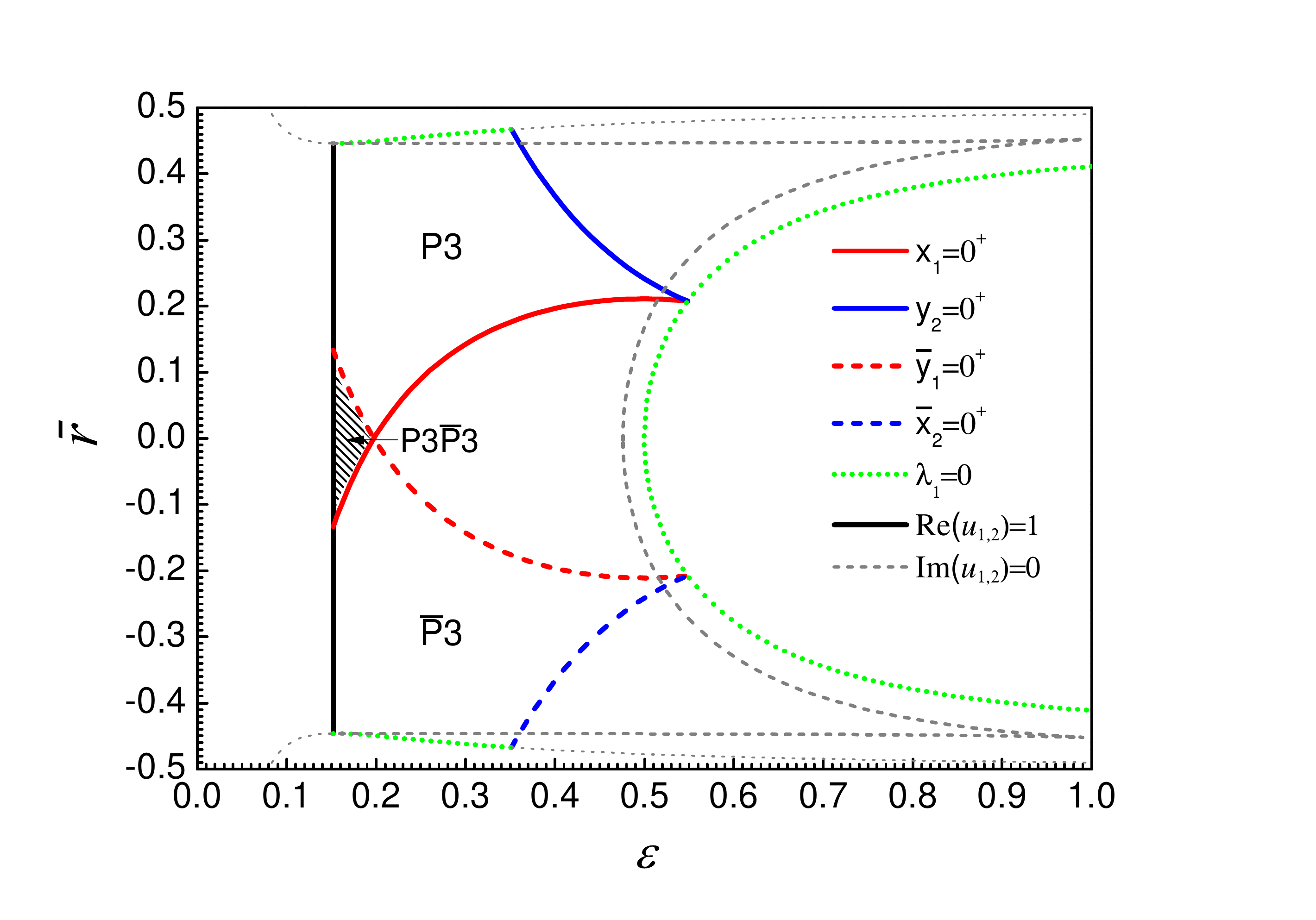}
\setlength{\abovecaptionskip}{0pt}
\setlength{\belowcaptionskip}{0pt}
\caption{(Color online)
{\bf Analysis of the period-3 attractor}. The blue and red curves represent
border collision bifurcations. The green dotted curves represent the critical
condition for the largest Lyapunov exponent $\lambda_1$. The black line
denotes the critical condition of the real part of complex conjugate
eigenvalues. The dash lines indicate zero imaginary part of the complex
conjugate eigenvalues.}
\label{fig:P3analysis}
\end{figure}

Our detailed calculation reveals that there are two types of border collision
bifurcations with the critical conditions given by
\begin{equation} \label{eq:P3BCBy}
A: \left\{
\begin{array}{c}
(x_2^*, y_2^*)=\mathbf{F}^3[(x_2^*, y_2^*)],\\
y_2^*=0^+,\\
\end{array}
\right.
\end{equation}
\begin{equation} \label{eq:P3BCBx}
B: \left\{
\begin{array}{c}
(x_1^*, y_1^*)=\mathbf{F}^3[(x_1^*, y_1^*)],\\
x_1^*=0^+,
\end{array}
\right.
\end{equation}
where the superscript ``+'' denotes the situation where the orbital point
collides with the discontinuous border from the positive side. The
stability condition of the period-3 attractor can then be obtained.
In particular, from Fig.~\ref{fig:P3orbits}, we have that the orbital
points of the attractor satisfy the conditions
$(x_1^*>0, y_1^*<0)$, $(x_2^*<0, y_2^*>0)$ and $(x_3^*>0, y_3^*>0)$.
The Jacobian matrix evaluated at the attractor is
\begin{equation} \label{eq:P3Jacobian}
\mathbf{DF}^3 = \mathbf{G}\!\left(\!\!
\begin{array}{cc}
\alpha & 0 \\
0 & \alpha \\
\end{array}\!\!
\right)\!
\mathbf{G}\!\left(\!\!
\begin{array}{cc}
\beta & 0 \\
0 & \alpha \\
\end{array}\!\!
\right)\!
\mathbf{G}\!\left(\!\!
\begin{array}{cc}
\alpha & 0 \\
0 & \beta \\
\end{array}\!\!
\right),
\end{equation}
with $\mathbf{G}$ being the coupling matrix
\begin{equation}
\left(
  \begin{array}{cc}\!
    1-\varepsilon(\frac{1}{2}-\bar{r}) & \varepsilon(\frac{1}{2}-\bar{r}) \\
    \varepsilon (\frac{1}{2}+\bar{r}) & 1-\varepsilon (\frac{1}{2}+\bar{r}) \\
  \end{array}\!
\right).
\end{equation}
From the characteristic equation Eq.~(\ref{eq:P3Jacobian}), we get
\begin{eqnarray} \label{eq:P3stable}
\Delta & = & (1-\varepsilon)^3\alpha^4\beta^2, \\ \nonumber
\tau & = & \frac{1}{4}\alpha(\varepsilon-2)
[((4\bar{r}^2-1)(\alpha-\beta)^2-4\alpha\beta)\varepsilon^2 \\ \nonumber
& - & 4\alpha\beta(\varepsilon+1)].
\end{eqnarray}
Combining Eqs.~(\ref{eq:RealLE})-(\ref{eq:P3BCBx}) and (\ref{eq:P3stable})
leads to the critical conditions for the period-3 attractor to be stable.

\begin{figure}[t]
\includegraphics[width=1.0\linewidth]{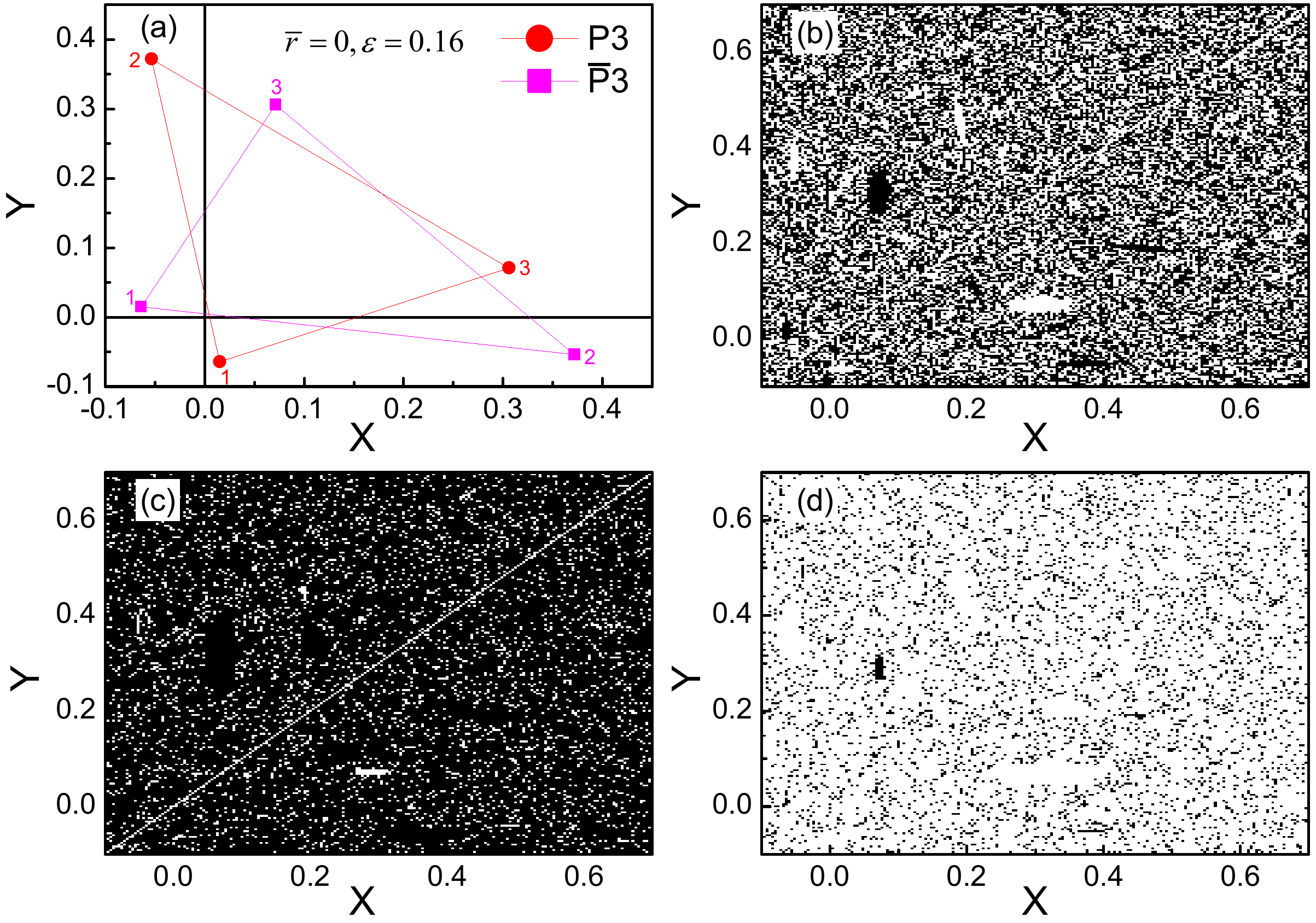}
\setlength{\abovecaptionskip}{0pt}
\setlength{\belowcaptionskip}{0pt}
\centering
\caption{(Color online)
{\bf Coexistence of two distinct period-3 attractors}. (a) Two types of
period-3 attractors that coexist in phase space. The basins of attraction of
the attractors for (b) $\bar{r}=0$ and $\varepsilon=0.16$, (c) $\bar{r}=0.05$
and $\varepsilon=0.16$, and (d) $\bar{r}=-0.05$ and $\varepsilon=0.16$, where
the dark and blank regions represent the basins of attraction of the
P3 and $\mathrm{\bar{P}}3$ attractors, respectively.}
\label{fig:P3basin}
\end{figure}

Figure~\ref{fig:P3analysis} shows the results from the stability analysis.
The stable period-3 attractor exists in the region surrounded by the curves of
stability (denoted by the green dotted curves and the black line) and border
collision bifurcations (denoted by red and blue curves). Comparing
Fig.~\ref{fig:P3analysis} with Fig.~\ref{fig:phasediagram}, we find a
good agreement between the theoretical analysis and the numerically
calculated structure of the parameter space for the period-3 attractor.
Specifically, for a fixed value of $\bar{r}$, as the coupling parameter
$\varepsilon$ is increased, the period-3 attractor undergoes a border
collision bifurcation before it becomes unstable, corresponding to the
the sudden transition from low-dimensional chaos to a periodic attractor, as
shown in Fig.~\ref{fig:Lypunovexp}. In addition, there is a region surrounded
by $x_1$, $\bar{y}_1$ and the stability curve, as marked by the oblique lines,
which explains the emergence of two types of period-3 attractors. Further
support for the coexistence of the two types of attractors can be obtained by
computing the basins of attraction, as shown in Fig.~\ref{fig:P3basin}. As
the bifurcation parameter is varied, the basins of the two types of period-3
attractors change.

\subsection{Occurrence of periodic attractors of period greater than three}
\label{subsec:HPA}

\begin{figure}[t]
\centering
\includegraphics[width=1.0\columnwidth]{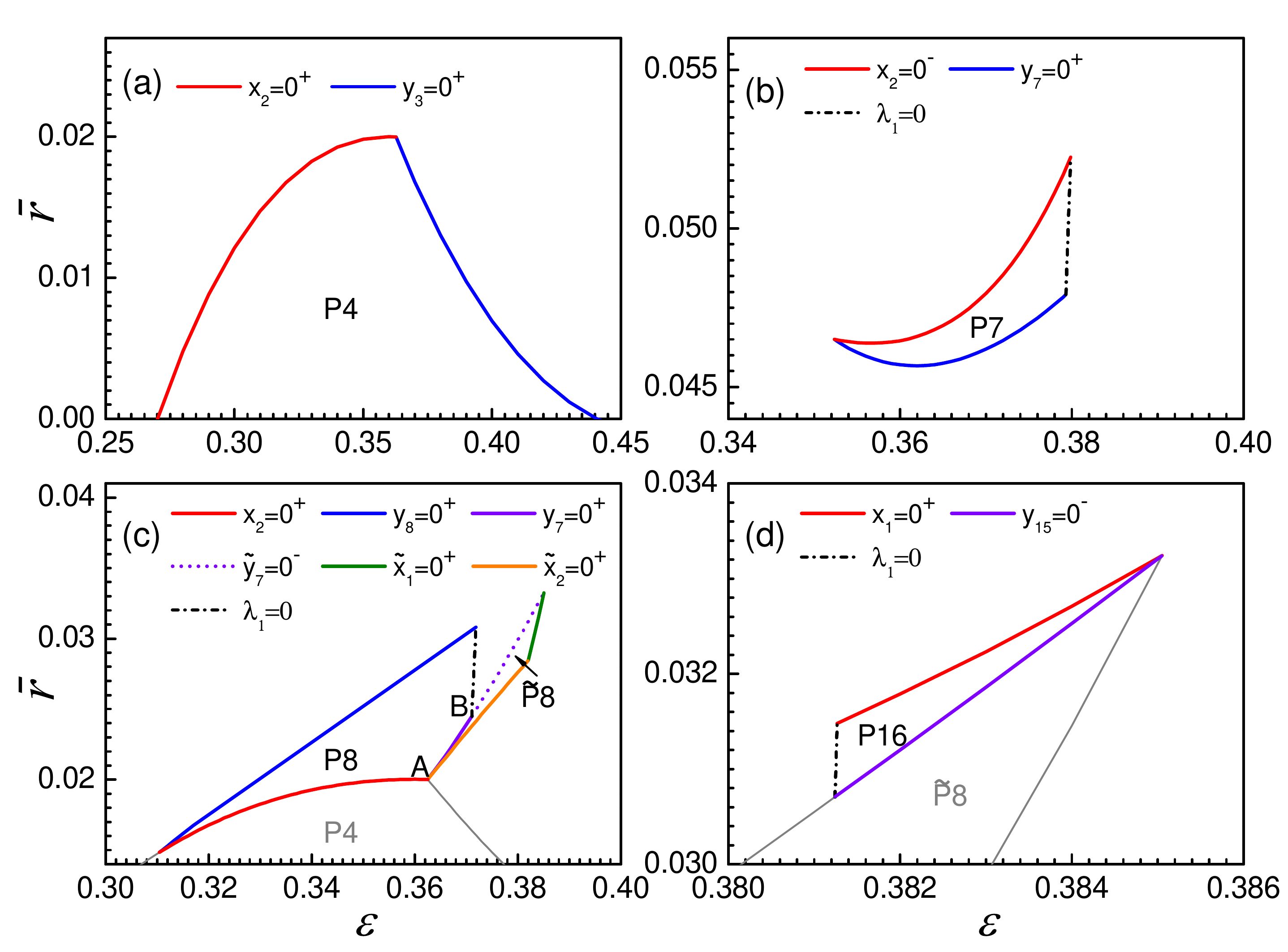}
\setlength{\abovecaptionskip}{0pt}
\setlength{\belowcaptionskip}{0pt}
\centering
\caption{(Color online)
{\bf Rise of periodic attractors of period greater than three}.
(a-d) Theoretically obtained stability regions for period-4, period-7,
period-8, and period-16 attractors, respectively.}
\label{fig:HPanalysis}
\end{figure}

Combining the linear stability and border collision bifurcation analyses,
we can obtain the existing conditions of periodic attractors of various
periods. Figure~\ref{fig:HPanalysis} shows the theoretical results for
periodic attractors of period-4, 7, 8, and 16 for $\bar{r}\geq 0$. We see
that, except for the period-4 attractor whose existing condition is
determined solely by border collision bifurcation, the emergence and
existence of periodic attractors of higher periods are due to the mixed
``action'' of stability and border collision bifurcation. We also find
border collision induced period-doubling bifurcations. For example, a
period-8 attractor (P8) arises after the period-4 orbit collides with
the discontinuous border, as shown in Fig.~\ref{fig:HPanalysis}(c), and
a periodic attractor of period-16 emerges after an alternative type of
period-8 attractor ($\mathrm{\tilde{P}}$8) collides with the border,
as shown in Fig.~\ref{fig:HPanalysis}(d). Further, the P8 and
$\mathrm{\tilde{P}}$8 attractors can convert into each other through
the collision that occurs on the $AB$ curve, as shown in
Fig.~\ref{fig:HPanalysis}(c). In general, as the period increases, the
area of the periodic attractor in the parameter space diminishes quickly.

\subsection{Globally coupled maps} \label{subsec:GCM}

\begin{figure}[t]
\centering
\includegraphics[width=1.0\linewidth]{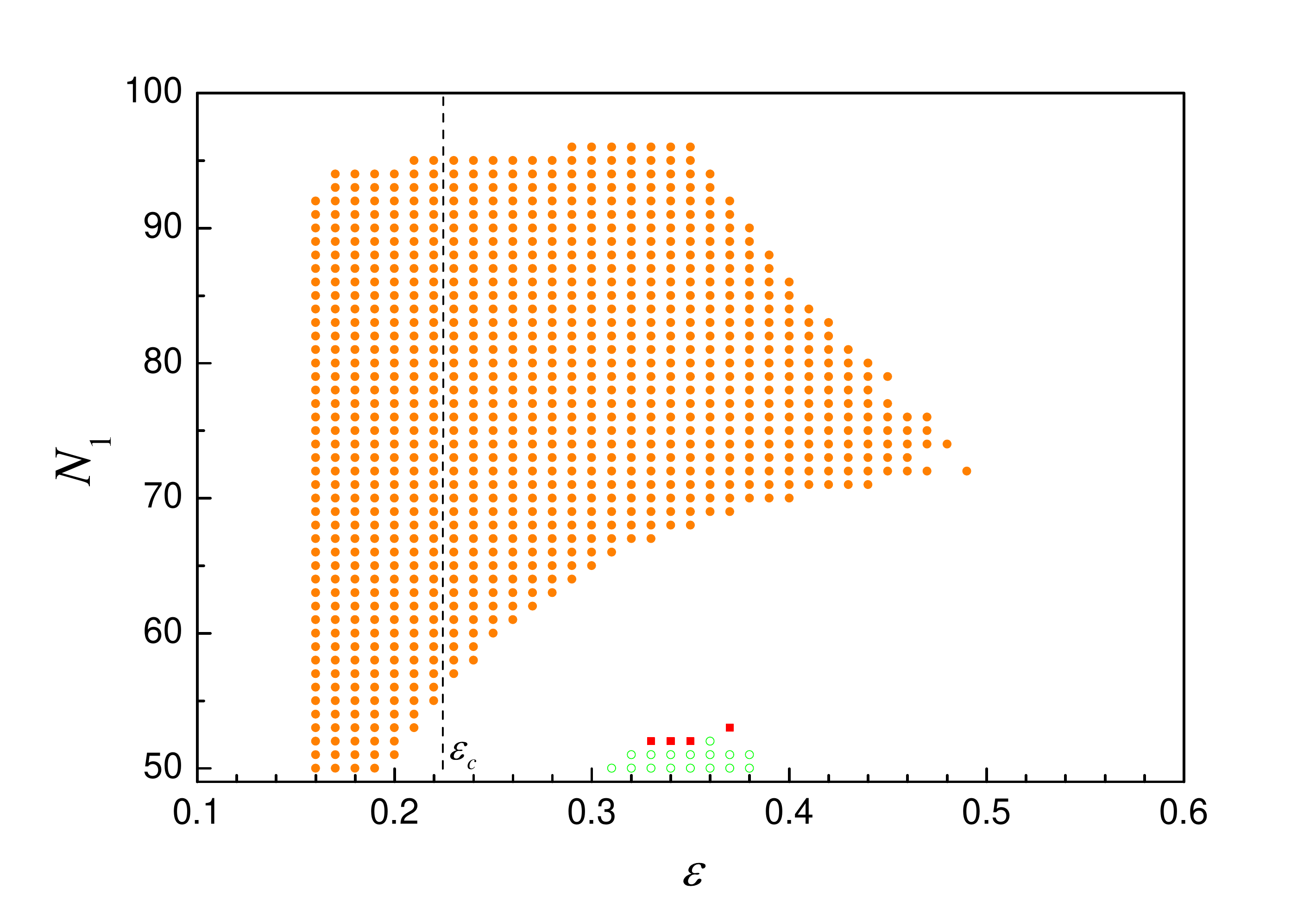}
\setlength{\abovecaptionskip}{0pt}
\setlength{\belowcaptionskip}{0pt}
\centering
\caption{(Color online)
{\bf Two-cluster state in a globally coupled nonsmooth map system}.
The size of the network is $N=100$ and $N_1$ represents the size of the
largest cluster. The two-cluster states of period-3, 4, and 8 are
represented by the orange solid, green open, and red square dots,
respectively. Each state is obtained using $10^5$ random initial
conditions.}
\label{fig:GCMtwoclusters}
\end{figure}

The occurrence of periodic attractors as a precursor of transition to
high-dimensional chaos in nonsmooth systems is a general phenomenon that
occurs in systems of globally coupled piecewise linear maps
[Eq.~(\ref{eq:GCM})]. For such a system, a variety of collective
dynamical states can arise. In particular, high-dimensional chaos manifests
itself as asynchronous chaos, whereas low-dimensional chaos corresponds to
globally synchronous chaos and, in the ``buffer'' region of periodic
attractors, periodic synchronization occurs. The parameter region in which
various two-cluster states occur is shown in Fig.~\ref{fig:GCMtwoclusters},
which qualitatively agrees with the phase diagram in
Fig.~\ref{fig:phasediagram}. Note that, not all stable two-cluster states
can be observed in a globally coupled system of finite size. In such
a system, multistability~\cite{FGHY:1996,FG:1997,KFG:1999,KF:2002,FG:2003,NFS:2011,Pateletal:2014,PF:2014,YHL:2016,LG:2017}
is common, and the
basin of attraction of a stable attractor can have a fractal structure,
on which small perturbations can have a significant effect. Certain
states are thus not physically observable. Note also that the result in
Fig.~\ref{fig:phasediagram} in fact corresponds to the thermodynamic limit
$N\to\infty$, but in Fig.~\ref{fig:GCMtwoclusters}, the network size $N$ is
finite. This leads to the small discrepancies between
Figs.~\ref{fig:GCMtwoclusters} and \ref{fig:phasediagram}.

\section{Discussion} \label{sec:discussion}

Historically, the discoveries of four distinct routes to low-dimensional
chaos with one positive Lyapunov exponent:
period-doubling~\cite{Feigenbaum:1978}, intermittency~\cite{PM:1980},
crisis~\cite{GOY:1983}, and quasiperiodicity~\cite{RT:1971,GS:1975,NRT:1978},
led to fundamental insights into and an understanding of the occurrence of
chaotic behaviors in natural systems and henceforth played an important role
in the development of nonlinear dynamics. Transition to high-dimensional chaos,
chaos with multiple positive Lyapunov exponents, has also been studied but
only for smooth dynamical
systems~\cite{HL:1999,HL:2000,DL:2000,PSM:2001,PM:2005}.
In such systems, a typical route to high-dimensional chaos is that the
second Lyapunov exponent passes through zero smoothly from the negative
side as a system parameter varies. The generality of this route lies in
regarding the underlying dynamical system as consisting of a number of
mutually interacting subsystems, some exhibiting low-dimensional
chaos. The chaotic subsystems then provide a kind of ``driving'' to
other subsystems. As a bifurcation parameter changes, an additional positive
Lyapunov exponent can arise. The nature of chaotic driving stipulates that
the second exponent becomes positive in a smooth
fashion~\cite{HL:1999,HL:2000,DL:2000}, a feature that is characteristic
of the transition to chaos in random dynamical
systems~\cite{YOC:1990,RHKK:2000,LLBS:2002,LLBS:2003}.

The main question addressed in this paper is whether transition to
high-dimensional chaos in nonsmooth dynamical systems can follow a
characteristically different route than that in smooth dynamical systems. The
answer is affirmative. In particular, using the paradigmatic setting of coupled
nonsmooth maps, we have uncovered a route in which a periodic attractor
arises as a precursor to high-dimensional chaos. That is, as a bifurcation
parameter is varied from the regime of a low-dimensional chaotic attractor,
an interval in which the attractor of the system is periodic occurs, after
which a high-dimensional chaotic attractor is born. In a two-dimensional
parameter space, the regions of low- and high-dimensional chaos are
separated by an open, ``bubble'' region of periodic attractors. As we
have shown, the route to high-dimensional chaos is characteristically
different from that in smooth dynamical systems, and the associated feature
in the parameter space is also distinct from that about the occurrence of
a periodic window (c.f., Fig.~\ref{fig:schematic}). Our analysis indicates
that the emergence of the ``bubble'' region can be attributed to border
collision bifurcations that occur commonly in nonsmooth dynamical systems.
Numerical computations have also revealed that there are parameter regions
in which a high-dimensional chaotic attractor can arise smoothly from a
low-dimensional one, as in smooth dynamical systems. The general finding
is then that, in nonsmooth dynamical systems, smooth and discontinuous
routes to high-dimensional chaos coexist in the parameter space. From the
perspective of transition to high-dimensional chaos, nonsmooth dynamical
systems thus offer richer behaviors than smooth dynamical systems.

\section*{Acknowledgments}

This work was supported by the National Natural Science Foundation of China
(Grant No.~11645005). YCL would like to acknowledge support from the Vannevar
Bush Faculty Fellowship program sponsored by the Basic Research Office of
the Assistant Secretary of Defense for Research and Engineering and
funded by the Office of Naval Research through Grant No.~N00014-16-1-2828.

%\bibliographystyle{elsarticle-num}
%\bibliographystyle{naturemag}
%\bibliographystyle{apsrev4-1}
%\bibliographystyle{ScienceAdvances}
%\bibliography{NSDS}

\end{document}